\documentclass[amsmath, amssymb, aps, prb, twocolumn, notitlepage, superscriptaddress]{revtex4-1}

\usepackage{graphicx}
\usepackage{dcolumn}
\usepackage{bm}
\usepackage[usenames]{color}
\usepackage{slashed}
\usepackage[utf8]{inputenc}
\usepackage[caption=false]{subfig}
\usepackage{hyperref}
\usepackage{enumitem}
\begin{document}
\title{Quantum aspects of “hydrodynamic” transport from weak electron-impurity scattering}
\author{Aaron Hui}
\affiliation{School of Applied \& Engineering Physics, Cornell University, Ithaca, New York 14853, USA}

\author{Samuel Lederer}
\affiliation{Department of Physics, Cornell University, Ithaca, New York 14853, USA}

\author{Vadim Oganesyan}
\affiliation{Department of Physics and Astronomy, College of Staten Island, CUNY, Staten Island, NY 10314, USA}
\affiliation{Physics Program and Initiative for the Theoretical Sciences,
The Graduate Center, CUNY, New York, NY 10016, USA}

\author{Eun-Ah Kim}
\affiliation{Department of Physics, Cornell University, Ithaca, New York 14853, USA}

\date{\today}

\begin{abstract}
Recent experimental observations of apparently hydrodynamic electronic transport have generated much excitement.
However, the understanding of the observed non-local transport (whirlpool) effects and parabolic (Poiseuille-like) current profiles has largely been motivated by a phenomenological analogy to classical fluids.
This is due to difficulty in incorporating strong correlations in quantum mechanical calculation of transport, which has been the primary angle for interpreting the apparently hydrodynamic transport. Here we demonstrate that even free fermion systems, in the presence of (inevitable) disorder, exhibit non-local conductivity effects such as those observed in experiment because of the fermionic system's long-range entangled nature.
On the basis of explicit calculations of the conductivity at finite wavevector, $\sigma({\bf q})$, for selected weakly disordered free fermion systems, we propose experimental strategies for demonstrating distinctive quantum effects in non-local transport at odds with the expectations of classical kinetic theory. 
Our results imply that the observation of whirlpools or other ``hydrodynamic" effects does not guarantee the dominance of electron-electron scattering over electron-impurity scattering.
\end{abstract}

\maketitle

{\it Introduction --}
Recent experimental reports of peculiar transport phenomena in ultraclean graphene\cite{Crossno2016, Bandurin2016, Kumar2017, Ku2019, Sulpizio2019} and other materials\cite{Moll2016,Gooth2018, Gusev2018} have generated much excitement regarding the role of hydrodynamic transport in these experiments. 
In the absence of microscopic understanding of the hydrodynamic transport of electrons, these experiments have been interpreted largely through analogy with classical fluids. 
Although parabolic velocity profiles\cite{Ku2019,Sulpizio2019} and whirlpools\cite{Bandurin2016} are familiar hydrodynamic phenomena in classical fluids, reliance on this analogy deprives us of an angle to learn the role of quantum mechanics in experiment. 
Most importantly, the question of the role of impurities, always present in materials, remains open although it has been clear that they complicate any analysis\cite{Andreev2011,Levitov2016}.

Modern interest in the hydrodynamic theory of electronic transport was motivated by a sore need for a theoretical framework to describe quantum critical transport in a regime dominated by electron-electron scattering.\cite{Damle1997, Son2007, Sachdev2009} 
Exotic possibilities have been predicted for graphene near the charge neutrality point,\cite{Hartnoll2007, Fritz2008, Foster2009, Muller2009, Torre2015, Levitov2016} and electron viscosity has been linked to the strange metal normal state of cuprate superconductors\cite{Davison2014, Lucas2015a, Lucas2017, Zaanen2019}.
However, a microscopic understanding of such hydrodynamic transport is challenging due to the inherent theoretical difficulty associated with the strongly correlated regime. Pioneering works used kinetic theory to calculate the shear viscosity for graphene\cite{Muller2009,Briskot2015,Principi2016} and for 2D Fermi liquids\cite{Ledwith2019a}, yielding non-trivial predictions.
However, as the role of (unavoidable) impurity scattering has primarily been treated phenomenologically via relaxation time approximations\cite{Conti1999, Torre2015, Levitov2016, Lucas2018, Sulpizio2019}, it has not been examined in microscopic detail.

In this paper, we evaluate the effects of impurity scattering, and identify signatures of the quantum nature of electrons, in the phenomena of whirlpool formation and parabolic current profiles. To do so, we explicitly calculate the non-local conductivity $\sigma({\mathbf q})$ for free electrons scattering off weak impurities.
In contrast to a classical Maxwell-Boltzmann distributed gas, in which the shear viscosity is independent of density\cite{Maxwell1860}, our principal result is that viscous effects have a distinctive dependence on carrier concentration.
This arises because Fermi statistics introduces a density-dependent velocity scale $v_F \sim \sqrt{n_e}$ (in 2D) and restricts scattering to the vicinity of the Fermi surface, so that scattering is determined by the density of states.
We map out experimental strategies to reveal the quantum nature near the bottom of band and in the vicinity of van Hove singularity. 

{\it Phenomenology and classical hydrodynamics --}
The phenomenological description of zero-frequency viscous transport\cite{Torre2015, Levitov2016} extends Drude theory by including the kinematic shear viscosity (i.e. coefficient of momentum diffusion) as 
\begin{align}
    \mathbf{E} =& A \left(\gamma - \nu \nabla^2\right)\mathbf{J}
    \label{eq:L-F phenom}
\end{align}
where $A$ is a dimensionful prefactor ($m/(n_e e^2)$ for Drude theory), $\gamma$ is the current scattering rate, and $\nu$ is the kinematic shear viscosity. 
This equation has a characteristic length scale $r_d \equiv \sqrt{\nu/\gamma}$, which we dub the viscosity length scale.
Note that in the limit of $\gamma \rightarrow 0$, Eq.~\ref{eq:L-F phenom} becomes a linearized Navier-Stokes equation (assuming $\mathbf{J}\propto\mathbf{p}$), with $\nu$ the usual fluid viscosity. \footnote{Although the definition of shear viscosity in the absence of momentum conservation is controversial, we take Eq.~\eqref{eq:L-F phenom} as a phenomenological definition of viscosity following Refs.\cite{Torre2015, Levitov2016}}  
Eq.~\eqref{eq:L-F phenom} amounts to a Taylor expansion in momentum of the usual Drude response (at zero frequency). 
Hence this equation applies to any system with current; it is agnostic to whether the system is classical or quantum.

\begin{figure}
    \centering
    \subfloat[][]{
        \includegraphics[width=.98\linewidth]{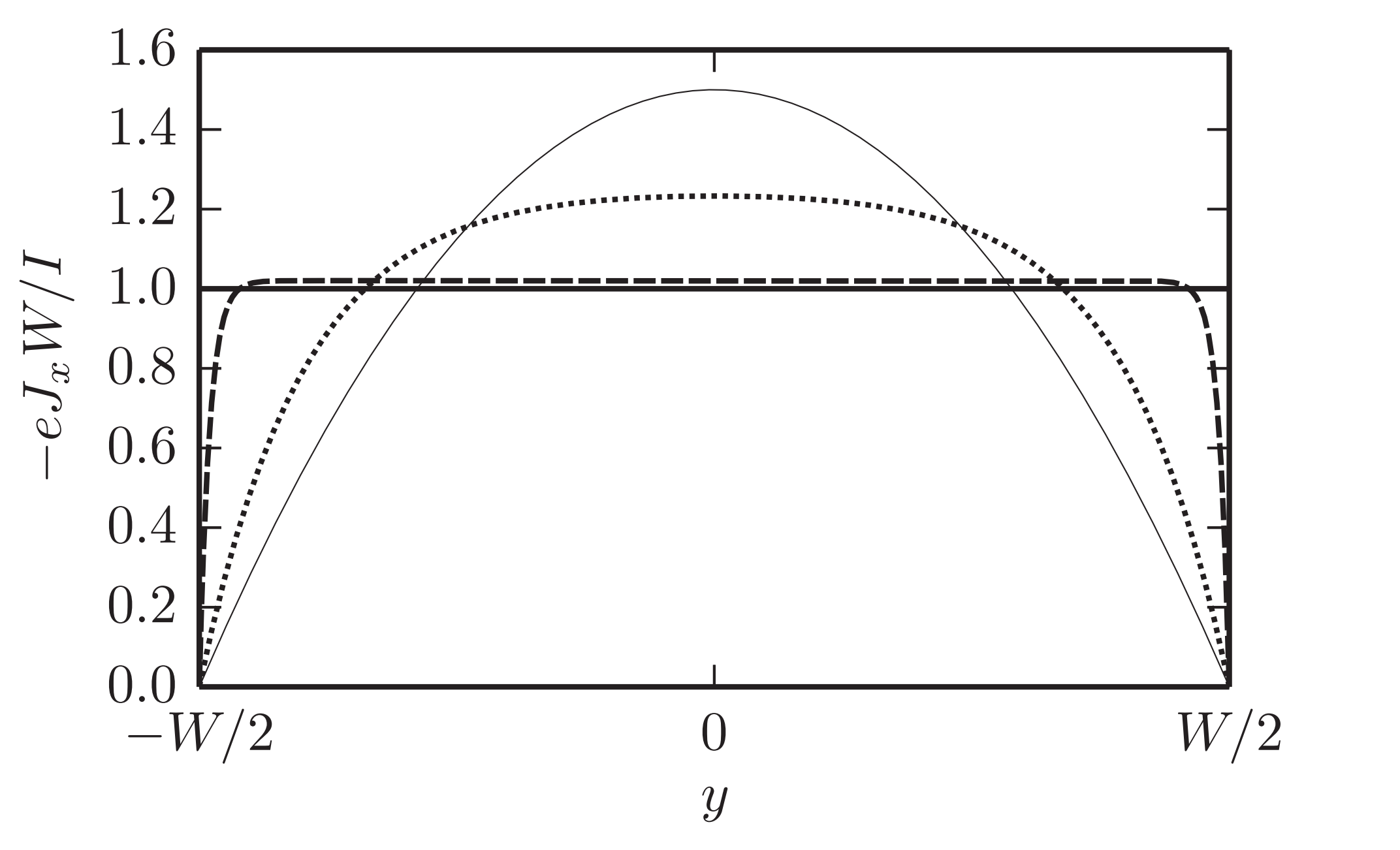}
        \label{fig:Torre Poiseuille}
    }
    
        \subfloat[][]{
        \includegraphics[width=.98\linewidth]{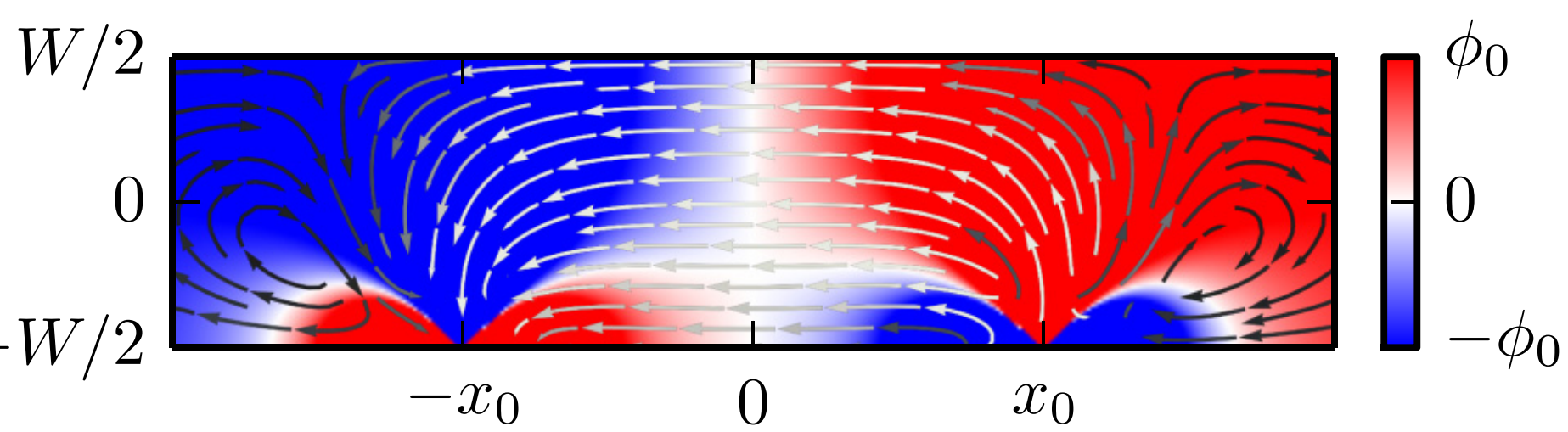}
        \label{fig:Torre Whirlpool}
    }
    \caption{Results from solving Eq.~\eqref{eq:L-F phenom} with no-slip boundary conditions, taken from Torre et al.\cite{Torre2015} \protect\subref{fig:Torre Poiseuille} A plot of the flow profile through a rectangular channel given by Eq.~\eqref{eq:Torre Poiseuille} for various values of $r_d/W$. For steady flow through a rectangular channel, the normalized current flow is rectangular for $r_d \ll W$ and parabolic for $r_d \gg W$. \protect\subref{fig:Torre Whirlpool} A heatmap of the potential $\phi$ and current streamlines for a current source and sink at $x_0$ and $-x_0$, respectively. White/black streamlines correspond to high/low current density. One finds that vortices form on the scale of $r_d$.}
    \label{fig:Torre}
\end{figure}

The existence of the length scale $r_d\equiv \sqrt{\nu/\gamma}$, associated with the kinematic shear viscosity $\nu$, immediately leads to the familiar hydrodynamic phenomena of parabolic current profiles and whirlpool formation. 
To see this, one can solve Eq.~\eqref{eq:L-F phenom} for the local current density $\mathbf{J}(\mathbf{r})$. 
For no-slip boundary conditions, the longitudinal flow down a rectangular channel of width $W$ is given by the formula\cite{Torre2015}
\begin{align}
    \frac{J_x(y) W}{I} = \left(1 - \frac{\cosh\frac{y}{r_d}}{\cosh \frac{W}{2r_d}}\right) \frac{1}{1 - \frac{2r_d}{W} \tanh\left(\frac{W}{2r_d}\right)}
    \label{eq:Torre Poiseuille}
\end{align}
As shown in Fig.~\ref{fig:Torre Poiseuille}, the flow profile is rectangular for $r_d \ll W$ and parabolic for $r_d \gg W$.
If one instead injects current laterally across the channel, as shown in Fig.~\ref{fig:Torre Whirlpool}, whirlpools of radius $\sim r_d$ will form.\cite{Torre2015, Levitov2016}

\begin{figure}
    \centering
    \subfloat[][]{
    \includegraphics[width=.5\linewidth]{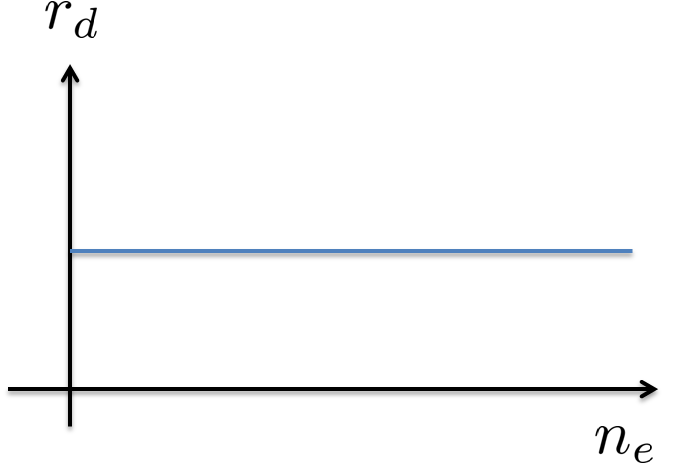}
    \label{fig:ideal gas}
    }
    \subfloat[][]{
    \includegraphics[width=.5\linewidth]{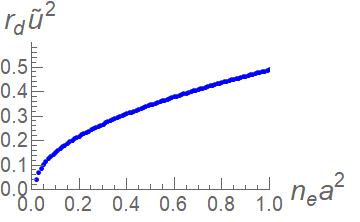}
    \label{fig:parabolic gas}
    }
    \caption{\protect\subref{fig:ideal gas} A plot of $r_d$ against (electron) gas density for the classical gas with impurities. The vortex radius is density-independent in this case. \protect\subref{fig:parabolic gas} The equivalent plot for a degenerate electron gas with a parabolic dispersion, at $u = 0.1 \frac{\hbar^2}{ma}$. We measure $r_d$ and $n_e$ in units of the lattice constant $a$ and $a^{-2}$, respectively, and introduced a dimensionless measure of disorder strength $\tilde{u} = \frac{u ma}{\hbar^2}$ so that the quantity $r_d \tilde{u}^2$ is independent of disorder strength.
    }
\end{figure}

For a 2D classical (Maxwell-Boltzmann) ideal gas of particles scattering off of dilute impurities, the velocity is set by temperature $T$ via the equipartition theorem as $v = \sqrt{2 k_B T/m_e}$. 
Since the mean free path is set by the cross section $\sigma_\text{imp}$ and the number density $n_\text{imp}$ of impurities as $l_\text{mfp} \sim 1/(n_\text{imp} \sigma_\text{imp})$,\footnote{This is slightly different from Maxwell's original model \cite{Maxwell1860} of rigid spheres, where $l_\text{mfp} \sim 1/(n_\text{gas}\sigma_\text{gas})$ since the collisions are with other gas particles.} the scattering rate is $\gamma=v/l_\text{mfp}$, independent of gas density. 
Moreover, it is known\cite{Lifshitz2012vol10} that the kinematic shear viscosity for weakly interacting classical gas is given by 
\begin{align}
    \nu \sim v l_\text{mfp}.
\end{align}
Hence in this classical system with impurities, the shear ``viscosity'' $\nu$ (phenomenologically defined in Eq.~\eqref{eq:L-F phenom}) and the vortex radius $r_d \sim l_\text{mfp}$ will be independent of the gas density as sketched in Fig.~\ref{fig:ideal gas}.

{\it Model and Formalism --}
The finite ${\mathbf q}$ conductivity $\sigma({\mathbf q})$ is related to the viscosity $\nu$ by inverting Eq.~\eqref{eq:L-F phenom}, which in the limit of small momenta gives
\begin{align}
    \mathbf{J} =& (\sigma_0 - \sigma_2\nabla^2) \mathbf{E}
    \label{eq:current expansion}
\end{align}
where $\sigma_0$ and $\sigma_2$ are the $\mathcal{O}(q^0)$ and $\mathcal{O}(q^2)$ pieces of $\sigma(q)$, respectively; the term linear in $q$ vanishes by inversion symmetry. 
These new parameters are related to the collision rate and viscosity of Eq.~\eqref{eq:L-F phenom} as $\sigma_0 = 1/(A\gamma)$ and $\sigma_2 = - \nu/(A\gamma^2)$.
In terms of $\sigma_0$ and $\sigma_2$,  the viscosity length scale $r_d$ is
\begin{align}
r_d \sim \sqrt{-\frac{\sigma_2}{\sigma_0}}
\label{eq:rd}
\end{align}
Of course, the conductivity $\sigma^{ij}$ is in actuality a rank-2 tensor, and hence $(\sigma_2)^{ij}_{\alpha\beta}$ is a rank-4 tensor. We have suppressed the tensor indices because the relevant components are parametrically equivalent,\footnote{There are subtleties regarding the formal equivalence between $\sigma_2$ and the shear viscosity $\nu$ which we are ignoring\cite{Bradlyn2012} in favor of the phenomenological definition of viscosity given by Eq.~\eqref{eq:L-F phenom}. Ultimately, we are interested in the experimental observable $r_d$, so the subtleties in the definition of viscosity do not pertain to us.} and will be using at $-(\sigma_2)_{xx}^{xx}/\sigma_0^{xx}$ as our estimate for $r_d^2$.
Often, transport calculations are done in the $q\rightarrow 0$ limit. However, obtaining non-local transport phenomena requires calculating at finite $\mathbf{q}$, in particular $\sigma_2 \propto \nu$.
The presence of finite $\mathbf{q}$ significantly complicates the calculations,\cite{Liu1970} as it breaks spatial symmetries and introduces angular dependencies in the integrand.

For our microscopic fermion model with weak impurity scattering, we consider $H = H_\text{kin} + H_\text{imp}$ with the kinetic term $H_\text{kin}$ and the impurity potential $H_\text{imp}$ given by
\begin{align}
    H_\text{kin} =& \frac{1}{\beta}\sum_{ik_n}\xi_\mathbf{k} c^\dagger_{\mathbf{k},ik_n} c_{\mathbf{k},ik_n}, \label{eq:Hkin}\\
    H_\text{imp} =& \frac{1}{\beta}\sum_{ik_n}\frac{1}{\beta} \sum_{iq_n} \int \frac{d^2 q}{(2\pi)^2} V(\mathbf{k}) c^\dagger_{\mathbf{k}^+,ik_n^+} c_{\mathbf{k}^-,ik_n^-}.
    \label{eq:Himp}
\end{align}
Here $\xi_\mathbf{k} = \epsilon_\mathbf{k} - \mu$ is the dispersion measured relative to the chemical potential, $(\mathbf{k}^\pm, ik_n^\pm) = (\mathbf{k} \pm \mathbf{q}/2, ik_n \pm iq_n/2)$ and $V(\mathbf{k})$ is the impurity potential in momentum space. 
We work in the $T\rightarrow 0$ limit.
For simplicity, we consider a Gaussian-distributed impurity potential where $\langle V(\mathbf{x}) \rangle = 0$ and $\langle V(\mathbf{x}) V(\mathbf{y}) \rangle = u^2 \delta(\mathbf{x}-\mathbf{y})$.
Thus, the disorder line transfers all momenta with equal weight $u^2$ but transfers no frequency. For the most part we will be content with only the perturbative treatment of disorder, which is expected to break down near band edges (dilute electrons or holes) and at the van Hove singularity. 

To calculate the conductivity, we use the Kubo formula
\begin{align}
    \sigma^{ij}(\mathbf{q},\omega+i0^+) = \frac{i}{\omega+i0^+}\left[\Pi^{ij}(\mathbf{q},\omega+i0^+) + \frac{n_e e ^2}{m}\delta^{ij}\right]
\end{align}
where $n_e$ is the average carrier density and $m$ is the particle mass\footnote{The mass generically has tensor structure which we have suppressed here for ease of presentation, as the diamagnetic piece will not play any significant role throughout this paper}. 
This requires us to calculate the current-current correlator $\Pi^{ij}$.
As we are interested in DC non-local response, we will be working in the limit $\omega\rightarrow 0$ and $v_F q \ll \gamma$, where $\gamma = -2 \operatorname{Im} \Sigma(\mathbf{q},\omega)$ is the scattering rate.\footnote{Although in general this limit requires a self-consistency check, for our disorder configuration $\Sigma$ is independent of $\mathbf{q}$, the regime always exists for sufficiently small $q$.}
We can separate contributions to $\Pi^{ij}$ into self-energy and vertex corrections; vertex corrections are negligible in this limit, as shown in Appendix~\ref{sec:vertex corrections}.
For the self-energy $\Sigma$, we will use first Born approximation\footnote{Recall that the $\mathcal{O}(u^1)$ piece amounts to a shift of the chemical potential $\mu$, and thus can be ignored.}
\begin{align}
    \Sigma(\mathbf{q},iq_n) = u^2 \int \frac{d^2 k}{(2\pi)^2} G_0(\mathbf{k},iq_n) 
\end{align}
where $G_{0}(\mathbf{q},iq_n) = (iq_n - \xi_\mathbf{q})^{-1}$ is the free Green's function. In addition, we will be ignoring the logarithmically UV divergent $\operatorname{Re} \Sigma$ by approximating it as a constant, in which case it amounts to a shift of $\mu$.
We also ignore the crossing diagrams and self-consistency diagrams of the self-energy.

Since we are only interested in dissipative response, using spectral function techniques we can rewrite the Kubo formula as
\begin{align}
    \operatorname{Re} &\sigma^{ij}(\mathbf{q},\omega) \nonumber\\
    &= \int_{-\omega}^{0} \frac{dx}{4\pi} \frac{d^2 k}{(2\pi)^2} \frac{A\left(\mathbf{k}^-,x\right) A\left(\mathbf{k}^+,x+\omega\right)}{-\omega} v^i(\mathbf{k}) v^j(\mathbf{k})
    \label{eq:Real Sigma}
\end{align}
where $A(\mathbf{k},\omega)$ is the spectral function and $v_i(\mathbf{k}) = \frac{\partial \epsilon_\mathbf{k}}{\partial k^i}$ is the current vertex factor (or velocity).\footnote{We assume that the diamagnetic term and the paramagnetic piece coming from $\operatorname{Im}1/(\omega + i\epsilon) \operatorname{Re} \protect \langle JJ \protect\rangle$ cancel.}
In 3D the relevant integrals can be evaluated via contour integration,\cite{Liu1970} but this approach cannot be extended to 2D. Hence we evaluate Eq.~\ref{eq:Real Sigma} numerically.
To obtain $\sigma_0$ and $\sigma_2$ as a function of carrier density $n_e$, for each fixed density we evaluate $\sigma^{ij}$ at fixed small $\omega$ ($= 10^{-9} \frac{\hbar}{ma^2} \approx 450$ KHz for a lattice constant $a = 5$\AA) for a number of momenta $qa \ll u^2 m^2 a^2/\hbar^2$ and perform a parabolic fit. 
For additional details, see the Appendix.

{\it Hydrodynamic transport and quantum effects --}
To target the manifestation of Fermi statistics through a density-dependent velocity, we consider a system with Fermi energy near the edge of a band. 
The dispersion is well approximated by the parabolic dispersion $\epsilon_\mathbf{k} = k^2/(2m)$.
The chemical potential $\mu$ is measured relative to the band bottom, i.e. $n_e = m \mu/(2\pi)$.
In this case, density of states is constant in 2D and the scattering rate $\gamma = -2 \operatorname{Im} \Sigma(\mathbf{q},\omega) = u^2 m$ is also a constant.
We use Eq.~\eqref{eq:Real Sigma} to evaluate $\operatorname{Re} \sigma^{ij}(\mathbf{q},\omega\rightarrow 0)$.
In our approach, $\sigma_0$ reproduces the known DC conductivity result $\sigma_0 =  \frac{n_e e^2}{m\gamma}$.
Extracting the viscosity length scale $r_d$ according to Eq.~\eqref{eq:rd}, we obtain the result shown in Fig.~\ref{fig:parabolic gas}, where we have plotted $r_d \tilde{u}^2$, where $\tilde{u} = \frac{uma}{\hbar^2}$ is the dimensionless disorder strength for lattice constant $a$.

The numerical results follow $r_d \sim \sqrt{n_e}$, as expected from the fact that the mean free path $l_\text{mfp}$ is the only length scale of our model and $l_\text{mfp} \sim v_F/\gamma \sim \sqrt{n_e}/(m \gamma)$. 
Such density dependence of the viscosity length scale is in clear contrast to the density-independent classical result of Fig.~\ref{fig:ideal gas}. 
For an experimental test of our prediction, the order of magnitude of $r_d$ needs to be experimentally accessible. 
The scale of $r_d$ will depend on the disorder strength in general, with $r_d \propto 1/u^2$ within the first Born approximation.
To obtain $r_d \approx 1 \mu$m, assuming $m$ is a free electron mass and $a \approx 5$\AA, we need $u \approx .02$ eV \AA.

We now turn to the effect of density of states on hydrodynamic transport. 
To see this effect in 2D, we propose tuning the Fermi level through the van Hove singularity. 
The recently developed experimental tuning parameters such as twist angle (in Moire systems\cite{Yan2012}) and uniaxial strain (in bulk crystals such as Sr$_2$RuO$_4$\cite{Barber2018}) could enable experimental tests of the proposal below.
For our calculation, we work in the limit where the impurity scattering rate is parametrically smaller than the distance $\delta \mu$ to the van Hove point, i.e. $\gamma \ll \delta \mu$, to have asymptotic control.
In the vicinity of a van Hove singularity, we consider the model Eq.~(\ref{eq:Hkin}-\ref{eq:Himp}) with the dispersion $\xi_\mathbf{k} = (k_x^2 - k_y^2)/(2m) - \delta \mu$, with $\delta \mu$ measuring the distance to the van Hove singularity.
This dispersion corresponds to considering only the vicinity of $(\pi,0)$ in the square lattice tight-binding model.
We regulate UV divergences in the continuum dispersion using a square cutoff $|k_x|, |k_y| < \Lambda$. 
Now the self-energy is given by 
\begin{align}
    \operatorname{Im} \Sigma(\mathbf{q}, \omega) = -\frac{m u^2}{2\pi} \operatorname{Re} \operatorname{coth}^{-1}\left(\frac{\Lambda}{\sqrt{-2m|\omega + \delta\mu| + \Lambda^2}}\right)
    \label{eq:Sigma-vH}
\end{align}
The logarithmic IR singularity at $\delta\mu = \omega = 0$ in the self-energy Eq.~\eqref{eq:Sigma-vH} captures the enhancement in impurity scattering due to the logarithmically diverging density of states near the van Hove singularity.

\begin{figure}
    \centering
    \includegraphics[width=.8\linewidth]{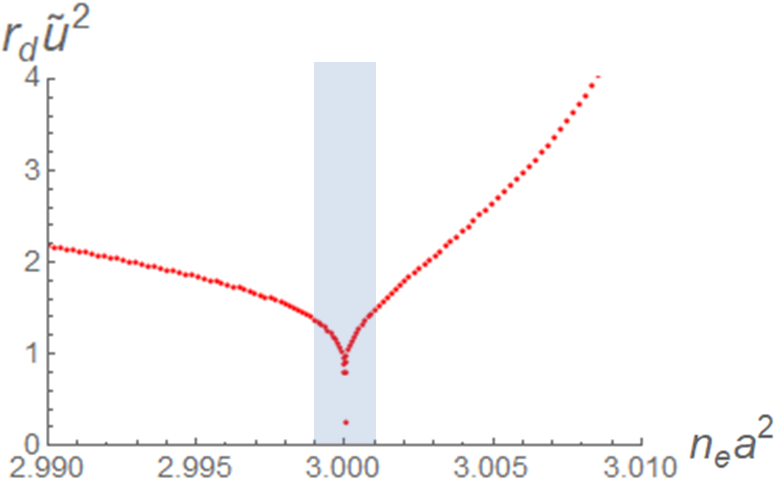}
    \caption{A plot of $r_d \tilde{u}^2$ against electron density for $u = 0.5 \frac{\hbar^2}{ma}$, where the Van Hove singularity is chosen to sit at $n_e a^2 = 3$. Notice that $r_d$ decreases on approach to the van Hove point due to the scattering enhancement from the logarithmically diverging density of states. The asymmetry about the van Hove point is a reflection of the anisotropy of the dispersion; we are only considering a single van Hove point corresponding to $(\pi,0)$ in a square lattice tight-binding model. The blue shaded region denotes the regime where $n_e - n_\text{vH} \ll \gamma$ and we expect self-consistent resummation of the self-energy to smooth out the singularity.}
    \label{fig:DOS}
\end{figure}

Fig.~\ref{fig:DOS} shows the computational results of the viscosity length scale $r_d$ in the vicinity of the van Hove singularity. 
To convert from $\delta \mu$ to $n_e - n_\text{vH}$, one uses the relation $n_e - n_\text{vH} = \int_0^{\delta \mu} \rho (x) |x| \, dx$, where $\rho(\mu)$ is the density of states as a function of chemical potential.
The singular suppression of $r_d$ reflects a diverging scattering rate as expected on the grounds of dimensional analysis: $r_d \sim v_F/\rm{Im}\Sigma$, so that $r_d \rightarrow 0$ as $\delta \mu \rightarrow 0$.
We expect an appropriate resummation of self-consistency diagrams to soften the singularity as impurity scattering blurs out the Fermi surface, and hence the van Hove point. 
This is expected of any van Hove effect in real systems.
Nevertheless, the suppression of the viscosity length scale $r_d$ is expected in the vicinity of the van Hove point. 
A confirmation of such suppression will be an unmistakable signature of a quantum effect. 

Recent experimental observations of the current flow profile in narrow channels~\cite{Ku2019,Sulpizio2019} and of negative non-local resistance from whirlpools~\cite{Bandurin2016} indicate that the above predictions can be tested. 
In particular, the ready tunability of Moire systems such as twisted bilayer graphene\cite{Li2010, Yan2012} would allow access to the carrier density dependence of the viscosity length scale $r_d\sim \sqrt{n_e}$ and the suppression of $r_d$ in the vicinity of a van Hove singularity. 

Finally, we comment on the finite frequency response, shown in Appendix~\ref{sec:frequency dependence}. 
An expansion of the finite frequency conductivity in the low frequency limit yields
\begin{align}
    \left|\frac{\sigma_2 (\omega)}{\sigma_0 (\omega)}\right| \approx r_d^2(1+B\omega^2).
\end{align}
Near the band edge, we find $r_d^2\sim v_F^2/\gamma^2$ and $B\sim 1/\gamma^2$, so $r_d^2/B \sim v_F^2$ is a disorder-independent quantity.
At frequencies $\omega \gtrsim \gamma$, the sign of $\sigma_2$ changes, signaling that the current oscillations are out of phase with the drive. 
For graphene, $\gamma$ has been estimated to be $650$ GHz.\cite{Bandurin2016}
In this regime, small finite momentum oscillations enhance rather than suppress the conductivity; we expect the formation of current stripes. 

{\it Summary and Discussion --}
To summarize, we considered hydrodynamic transport in a microscopic model of electrons under weak impurity scattering. 
The motivation was two-fold: (1) to study the effect of disorder and (2) to reveal quantum aspects. 
We have shown that apparently hydrodynamic phenomena such as formation of a parabolic current profile and a whirlpool can be caused entirely by weak disorder scattering. 
For this, we have explicitly calculated the viscosity length scale $r_d$, which sets the whirlpool size and the curvature of the current flow profile, by calculating the non-local conductivity $\sigma(\mathbf{q})$ and expanding it in powers of $q$.
Furthermore, we proposed experimental strategies to access quantum aspects of such transport phenomena by tracking carrier density dependence of $r_d$ and tuning to the vicinity of a van Hove point. These distinctly quantum signatures arise due to the long-range entangled nature of the free fermion system (i.e. its statistics).

Our results raise the question of how to distinguish impurity scattering effects from electron-electron interaction effects in experiments exhibiting hydrodynamic transport, namely parabolic current profile and whirlpool formation, also raised in Ref.~\onlinecite{Sulpizio2019}.
Indeed, viscosity itself needs to be carefully defined in the presence of impurities as momentum conservation is violated; finite $\mathbf{q}$ conductivity and the stress-strain correlator, both of which give viscosity in the clean limit,\cite{Bradlyn2012} are not necessarily linked in a dirty system.\cite{Burmistrov2019}
The role of impurity scattering in other hydrodynamic transport phenomena such as unusual temperature dependence of charge transport such as the Gurzhi effect~\cite{deJong1995,Kumar2017}, thermal transport anomalies~\cite{Crossno2016,Gooth2018}, and magnetotransport~\cite{Moll2016} will be topics of future theoretical studies. 
Here we focused on delta-function correlated disorder; finite-range disorder would introduce a new length scale, and it would be interesting to understand the influence of this length scale on $r_d$ and other transport phenomena.
Our results open doors to considering other forms of scattering, including electron-phonon and umklapp scattering in the future.
Another interesting future direction is the nature of the boundary, which is known to play an important role in determining viscous transport\cite{Kiselev2019}, in the weakly disordered regime.
Last but not least, it would be interesting to revisit ultraclean two-dimensional electron gases~\cite{deJong1995} to test our predictions of density dependence of $r_d$. 

\vspace{5mm}
{\noindent\bf Acknowledgements}
We thank Philip Kim, Leonid Levitov, Philip Moll, Andy Lucas, Srinivas Raghu, Jeevak Parpia, Subir Sachdev, Joerg Schmalian, Amir Yacoby, and Jan Zaanen for helpful discussions. A.H. was supported by the National Science Foundation Graduate Research Fellowship under Grant No. DGE-1650441 and by the W.M. Keck Foundation. SL was supported by a Bethe/KIC fellowship and by the W.M. Keck Foundation. VO was supported under DMR Grant No. 1508538. E-AK was supported by the W.M. Keck Foundation.

\bibliography{biblio}

\appendix
\onecolumngrid

\section{Feynman Rules}
The Feynman rules for our model are the following:

\begin{center}
\includegraphics[width = .45\linewidth]{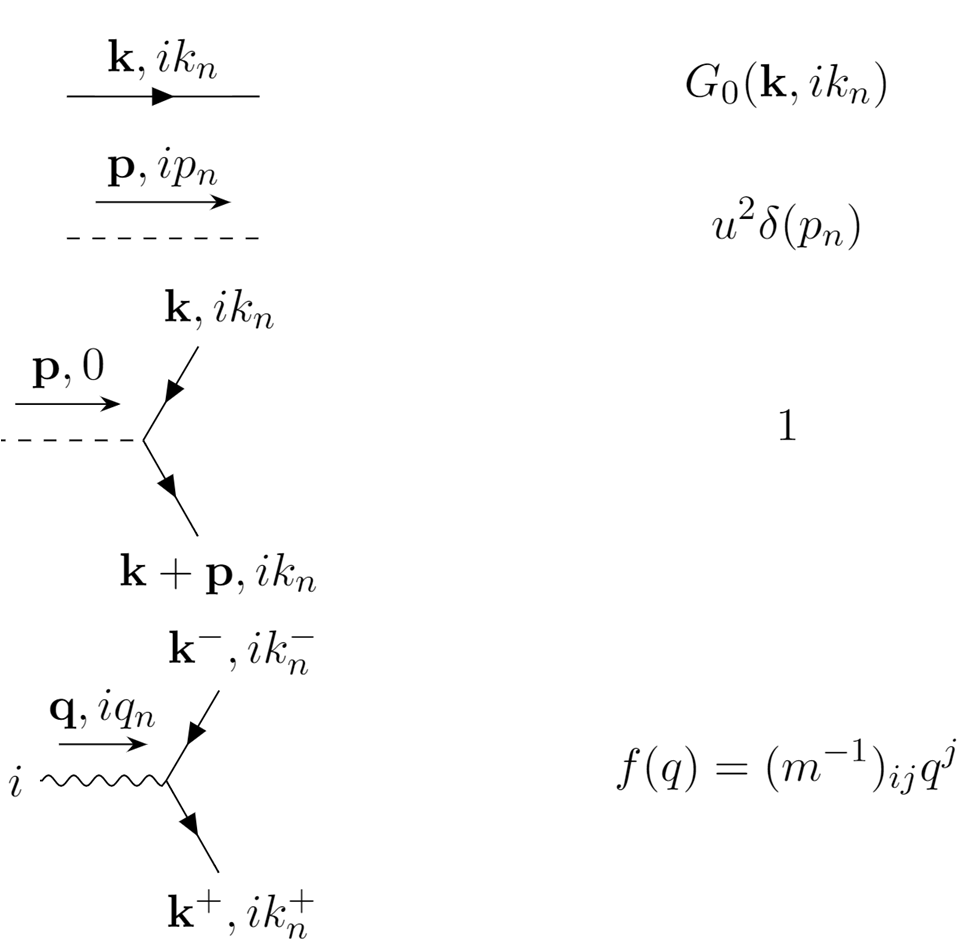}
\end{center}

\noindent where we've defined $(\mathbf{k}^\pm, ik_n^\pm) \equiv (\mathbf{k} \pm \frac{\mathbf{q}}{2}, ik_n \pm \frac{iq_n}{2})$.
The solid line corresponds to the free electron propagator $G_0 (\mathbf{k},ik_n) = \frac{1}{ik_n - \xi_\mathbf{k}}$.
The dashed line corresponds to the impurity interaction, which transfers all momenta but no frequency, and is momentum independent.
The impurity scattering vertex is just unit; as noted it transfers momenta but no frequency.
The current vertex, with an external photon line with polarization $i$, has a current vertex factor corresponding to velocity. 

\section{Kubo Formula: Spectral Function}

\begin{figure}
    \centering
    \subfloat[][Current-current correlator without vertex corrections]{
    \includegraphics[width=.3\linewidth]{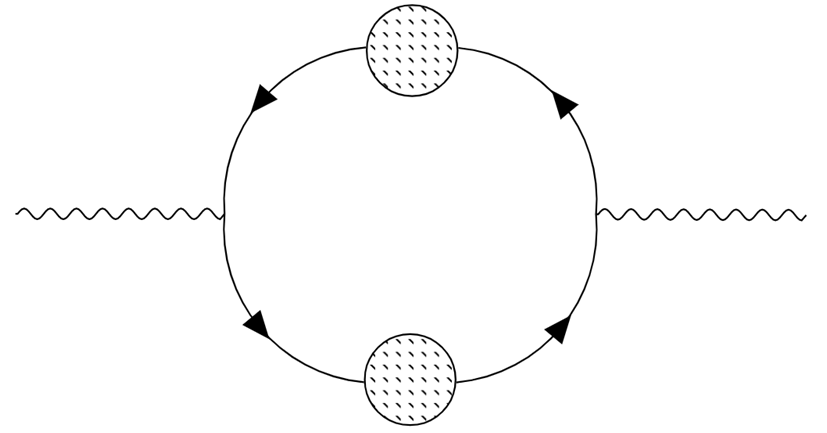}
    \label{fig: JJ correlator}
    }
    \subfloat[][The first Born approximation of the self-energy]{
    \includegraphics[width=.3\linewidth]{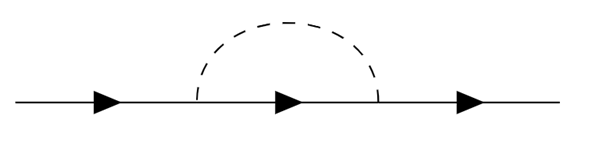}
    \label{fig:self-energy}
    }
    \subfloat[][Lowest order vertex correction]{
    \includegraphics[width=.3\linewidth]{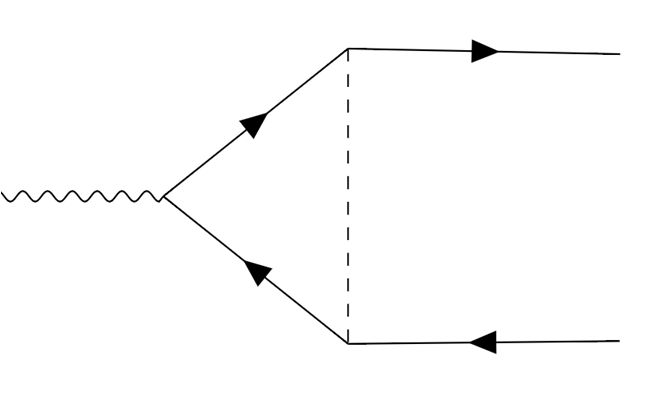}
    \label{fig:vertex correction}
    }
    \caption{Feynman diagrams}
\end{figure}

Calculating the current-current correlator involves evaluating diagrams of the form shown in Fig.~\ref{fig: JJ correlator}.
In the regime of interest of this paper, namely $\omega \rightarrow 0$, vertex corrections can be neglected at $q^2$ order in the conductivity $\sigma$, as shown in Appendix~\ref{sec:vertex corrections}.
Therefore, all that remains are self-energy corrections to the fermion propagator.

When $G(\mathbf{k},ik_n)$ has self-energy corrections, i.e. $G^{-1}(\mathbf{k},ik_n) = ik_n - \xi_\mathbf{k} -\Sigma(\mathbf{k},ik_n)$, branch cuts pose complications if one wants to perform Matsubara sums via contour integration. To get around this issue, we use the spectral function approach, which relies on the identity:

\begin{align}
G(\mathbf{k},ik_n) =& \int \frac{dx}{2\pi} \frac{A(\mathbf{k},x)}{ik_n - x}
\\
A(\mathbf{k},\omega) =& \frac{-2\operatorname{Im}\Sigma(\mathbf{k},\omega)}{\left[\omega-\xi_\mathbf{k}-\operatorname{Re}\Sigma(\mathbf{k},\omega)\right]^2 + \left[\operatorname{Im}\Sigma(\mathbf{k},\omega)\right]^2}
\end{align}
where $A(\mathbf{k},\omega) \equiv -2\operatorname{Im} G(\mathbf{k},\omega)$ is called the spectral function. It is a fact that $A(\mathbf{k},\omega) \geq 0$.\cite{Mahan2000book} 
This identity allows us to perform the Matsubara sum, moving the difficulties of evaluation to the integration.
We define $\mathbf{k}^\pm \equiv \mathbf{k} \pm \frac{\mathbf{q}}{2}$ for ease of presentation.

\begin{align}
\Pi_{\alpha\beta}(\mathbf{q}) =& (-1) \int\frac{d^2 k}{(2\pi)^2} \frac{dx dy}{(2\pi)^2} A\left(\mathbf{k}^-,x\right) A\left(\mathbf{k}^+,y\right) \frac{n_F(x) - n_F(y)}{iq_n + x - y} v_\alpha(\mathbf{k}) v_\beta(\mathbf{k})
\\
\operatorname{Im} \Pi_{\alpha\beta}(\mathbf{q},\omega) =& (-1) \int\frac{d^2 k}{(2\pi)^2} \frac{dx dy}{(2\pi)^2} A\left(\mathbf{k}^-,x\right) A\left(\mathbf{k}^+,y\right) \Big[n_F(x) - n_F(y)\Big] (-\pi) \delta(\omega + x - y)v_\alpha(\mathbf{k}) v_\beta(\mathbf{k})
\\
=& \int\frac{d^2 k}{(2\pi)^2} \frac{dx}{4\pi} A\left(\mathbf{k}^-,x\right) A\left(\mathbf{k}^+,x+\omega\right) \Big[n_F(x) - n_F(x+\omega)\Big]v_\alpha(\mathbf{k}) v_\beta(\mathbf{k})
\end{align}
In these equations, we suppressed $i0^+$ in the frequency, as we don't expect this to play any role due to the presence of an non-zero imaginary self-energy.

\begin{figure}
    \centering
    \includegraphics[width=.4\linewidth]{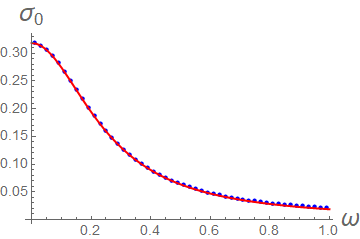}
    \caption{A plot of the zero-momentum conductivity $\sigma_0(\omega)$ for the fermion with parabolic dispersion, for $u=.1 \frac{\hbar^2}{ma}$. The blue points are numerical data, and the red line is not a fit, but the function $\frac{k_F^2}{4\pi m} \frac{\gamma}{\omega^2 + \gamma^2}$.}
    \label{fig:drude}
\end{figure}

To verify this is correct, for the fermion with parabolic dispersion we plotted the zero-momentum conductivity $\sigma_0(\omega)$ and find that it matches precisely with $\sigma_0(\omega) = \frac{k_F^2}{4\pi m} \frac{\gamma}{\omega^2 + \gamma^2}$, as shown in Fig.~\ref{fig:drude}. This corroborates our Drude theory expectations and that $\sigma_0 = \frac{k_F^2 e^2}{4\pi m \gamma} = \frac{n_e e^2}{m \gamma}$ as stated in the main text.

\section{Self-Energy}

In the model as stated in the main text, we need to evaluate the integral
\begin{align}
    \Sigma(\mathbf{q},iq_n) = u^2 \int \frac{d^2 k}{(2\pi)^2} G_0(\mathbf{k},iq_n) 
\end{align}
corresponding to the diagram shown in Fig.~\ref{fig:self-energy}.

\subsection{Parabolic Fermion}
The dispersion for the parabolic (spinless) fermion is given by $\xi_\mathbf{k} = k^2/(2m) - \mu$. We recall that the 2D density of states for this case is $m/(2\pi)$. 
\begin{align}
    \Sigma(\mathbf{q},iq_n) =& u^2 \int \frac{d^2 k}{(2\pi)^2} \frac{1}{ik_n - \epsilon_\mathbf{k} + \mu} \\
    =& u^2 \int d\epsilon_k \frac{m}{2\pi} \frac{1}{ik_n - \epsilon_\mathbf{k} + \mu} \\
    \Sigma(\mathbf{q},\omega + i0^+) =& u^2 \frac{m}{2\pi} \int_0^{\Lambda^2/(2m)} d\epsilon_\mathbf{k} P\frac{1}{\omega + \mu - \epsilon_\mathbf{k}} - i\pi \delta(\omega + \mu - \epsilon_\mathbf{k}) \\ 
    =& u^2 \frac{m}{2\pi} \ln \left(\frac{\Lambda^2/(2m)}{\omega + \mu} - 1\right) - i u^2 \frac{m}{2}
\end{align}
where $P$ denotes the principal value and we take a spherically symmetric cutoff $0<k<\Lambda$.
We find that the real part is logarithmically UV divergent, and the imaginary part is constant.

\subsection{Van Hove Fermion}
The dispersion for van Hove fermion is given by $\xi_\mathbf{k} = (k_x^2-k_y^2)/(2m) - \delta \mu$. We take cutoffs $-\Lambda < k_x,k_y < \Lambda$. As noted in the main text, and similar to the parabolic fermion, we ignore $\operatorname{Re} \Sigma$.
\begin{align}
    \operatorname{Im} \Sigma(\mathbf{q},\omega) =& -u^2 \pi \int \frac{d^2 k}{(2\pi)^2} \delta(\omega + \delta\mu - \epsilon_\mathbf{k}) \\ 
    =& -\frac{m}{2\pi} u^2 \operatorname{Re} \coth^{-1}\left(\frac{\Lambda}{\sqrt{-2m|\omega + \delta\mu| + \Lambda^2}}\right)
\end{align}

\section{Vertex Corrections}
\label{sec:vertex corrections}
In this section, we consider the lowest order vertex correction diagram, shown in Fig.~\ref{fig:vertex correction}, and show that the $q^2$ contribution to the conductivity $\sigma$ must vanish in the limit of $\omega \rightarrow 0$. We show this in two ways.

\subsection{Vertex corrections vanish as $\omega \rightarrow 0$}
We define $\mathbf{k}^\pm, ik_n^\pm \equiv \mathbf{k}\pm\frac{\mathbf{q}}{2}, ik_n \pm \frac{q_n}{2}$ and take a dispersion such that $\epsilon_\mathbf{k} = \epsilon_\mathbf{-k}$. This even-parity condition is satisfied for both the parabolic and van Hove dispersions. Recall that for impurity scattering, the disorder line transfers momenta but no frequency; since the disorder line (and vertex) is momentum-independent, the amputated vertex $\Gamma^i(\mathbf{q},iq_n; ik_n)$ is independent of the external fermion momentum $\mathbf{k}$.

\begin{align}
    \Gamma^i (\mathbf{q},iq_n; ik_n) =& u^2 \int \frac{d^2 k}{(2\pi)^2} G(\mathbf{k}^+,ik_n^+) G(\mathbf{k}^-, ik_n^-) k^i \\
    =& u^2 \int \frac{d^2 k}{(2\pi)^2} \frac{1}{ik_n^+ - \epsilon_{\mathbf{k}^+} - \Sigma(\mathbf{k}^+,ik_n^+)} \frac{1}{ik_n^- -\epsilon_{\mathbf{k}^-} - \Sigma(\mathbf{k}^-,ik_n^-)} k^i \\
    =& u^2 \int \frac{d^2 k}{(2\pi)^2} \frac{1}{iq_n - \epsilon_{\mathbf{k}^+} + \epsilon_{\mathbf{k}^-} - \Sigma(\mathbf{k}^+,ik_n^+) + \Sigma(\mathbf{k}^-,ik_n^-)} \left[\frac{1}{ik_n^- - \epsilon_{\mathbf{k}^-}} - \frac{1}{ik_n^+ - \epsilon_{\mathbf{k}^+}}\right] k^i \\
    \Gamma^i (\mathbf{q},\omega + i\epsilon; ik_n) =& u^2 \int \frac{d^2 k}{(2\pi)^2} \frac{1}{\omega + i\epsilon - \epsilon_{\mathbf{k}^+} + \epsilon_{\mathbf{k}^-} - \Sigma(\mathbf{k}^+,ik_n^+) + \Sigma(\mathbf{k}^-,ik_n^-)} \left[\frac{1}{ik_n^- - \epsilon_{\mathbf{k}^-}} - \frac{1}{ik_n^+ - \epsilon_{\mathbf{k}^+}}\right] k^i
\end{align}
In the second to last line we have decomposed via partial fractions. This is valid as long as the two fractions are never equal to each other (at finite $\mathbf{q}$).

We are interested in the $\omega \rightarrow 0$ limit, so we take $iq_n \rightarrow \omega + i\epsilon$ and set $\omega = 0$. \footnote{Formally $\epsilon \rightarrow 0$ first before anything, but we believe this order is fine since it introduces no divergences.} In this limit, $ik_n^\pm = ik_n \pm i\epsilon$.

Because we are considering a momentum-independent disorder strength, the self-energy cannot depend on momentum, i.e. $\Sigma(\mathbf{k},ik_n) = \Sigma(ik_n)$. We will also take the assumption that $\displaystyle \lim_{\omega\rightarrow 0}\Sigma(ik_n^+) = \lim_{\omega\rightarrow 0} \Sigma(ik_n^-)$.\footnote{For the free fermion, this is trivially true as $\Sigma = 0$. If one works in the first Born approximation, $\Sigma(ik_n) = i\gamma \operatorname{sign} (k_n)$, which also satisfies this condition as $k_n \neq 0$ for any finite $T$}

Because we are working at finite temperature and $\omega\rightarrow 0$, we have $ik_n^\pm - \epsilon_{\mathbf{k}^\pm} = ik_n - \epsilon_{\mathbf{k}^\pm}$, as we take $\epsilon \rightarrow 0$ before $T\rightarrow 0$. 

Putting this all together, we have
\begin{align}
    \Gamma^i (\mathbf{q},\omega + i\epsilon; ik_n) =& u^2 \int \frac{d^2 k}{(2\pi)^2} \frac{1}{i\epsilon - \epsilon_{\mathbf{k}^+} + \epsilon_{\mathbf{k}^-}} \left[\frac{1}{ik_n - \epsilon_{\mathbf{k}^-}} - \frac{1}{ik_n - \epsilon_{\mathbf{k}^+}}\right] k^i \\
    =& u^2 \int \frac{d^2 k}{(2\pi)^2} \left(P\frac{1}{- \epsilon_{\mathbf{k}^+} + \epsilon_{\mathbf{k}^-}} - i\pi\delta(-\epsilon_{\mathbf{k}^+} + \epsilon_{\mathbf{k}^-})\right) \left[\frac{1}{ik_n - \epsilon_{\mathbf{k}^-}} - \frac{1}{ik_n - \epsilon_{\mathbf{k}^+}}\right] k^i
\end{align}
where $P$ denotes the principal value.

It is immediately clear that the imaginary part vanishes identically due to the delta function. For the real part, consider the momentum inversion $\mathbf{k}\rightarrow -\mathbf{k}$ in the integrand. This sends $\epsilon_{\mathbf{k}^{\pm}} \rightarrow \epsilon_{\mathbf{k}^\mp}$ so that the integrand is odd under momentum inversion. Because of this, the real part must also vanish. Hence, $\Gamma^i$ is identically zero.

Assuming that $\Gamma^i$ is regular in $\omega$, this implies that $\Gamma^i$ is $\mathcal{O}(\omega)$ so that $\sigma(\mathbf{q},\omega)$ is also $\mathcal{O}(\omega)$. 

\subsection{The $q^2$ component of $\sigma$ is purely reactive}
Alternatively, we will show that the dissipative $q^2$ component of $\sigma$, i.e. $\sigma_2$, is zero. We first Taylor expand in $q$. 
\begin{align}
    \Gamma^i(\mathbf{q},iq_n; ik_n) =& u^2 \int \frac{d^2 k}{(2\pi)^2} G(\mathbf{k},ik_n^+) G(\mathbf{k},ik_n^-) k^i + \left[\partial_{k^\alpha} G(\mathbf{k},ik_n^+) G(\mathbf{k},ik_n^-) - G(\mathbf{k},ik_n^+) \partial_{k^\alpha} G(\mathbf{k},ik_n^-)\right] q^\alpha k^i \\
    =& u^2 \int \frac{d^2 k}{(2\pi)^2} \left[\partial_{k^\alpha} G(\mathbf{k},ik_n^+) G(\mathbf{k},ik_n^-) - G(\mathbf{k},ik_n^+) \partial_{k^\alpha} G(\mathbf{k},ik_n^-)\right] q^\alpha k^i \\
\end{align}

Notice that if we Taylor expand in $\omega$, the $\mathcal{O}(\omega^0)$ term vanishes, so that $\Gamma^i \sim \mathcal{O}(\omega)$. This implies that the $q^2$ component of the current-current correlator is $\mathcal{O}(\omega^2)$. However, we know that dissipative response functions, i.e. the current-current correlator, must be odd in frequency, hence for $\omega \rightarrow 0$ the $q^2$ component is purely reactive. Therefore, we know that $\sigma_2$ vanishes in the limit $\omega \rightarrow 0$.

\section{Frequency Dependence}
\label{sec:frequency dependence}
\begin{figure}
    \centering
    \subfloat[][]{
    \includegraphics[width=.4\linewidth]{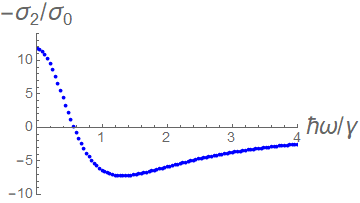}
    \label{fig:frequency}
    }
    \subfloat[][]{
    \includegraphics[width=.4\linewidth]{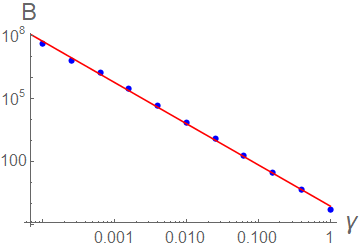}
    \label{fig:B disorder dependence}
    }
    \caption{\protect\subref{fig:frequency} A plot of $-\sigma_2/\sigma_0 \sim r_d^2$ against frequency at $u = .5 \frac{\hbar^2}{ma}$, normalized against the scattering rate $\gamma = u^2 m$. Around $\omega \sim \gamma/2$, the sign of $-\sigma_2/\sigma_0$ changes. \protect\subref{fig:B disorder dependence} A log-log plot of the $\gamma$ dependence of $B$, where the blue points are numerical data and the red line is a linear fit. We find $B \propto \gamma^{-2}$.}
\end{figure}

We remark on frequency-dependent behavior in the electron with parabolic dispersion. 
These characteristics also appear in the van Hove fermion as well.
In Fig.~\ref{fig:frequency}, we see that $r_d$ changes from positive to negative when $\omega \approx \gamma$.
As this corresponds to the fact that the current-current correlator changes sign at high frequency, this sign change is a reflection of the fact that the current will go out of phase with the drive.
In Fig.~\ref{fig:B disorder dependence}, we see that for $\frac{\sigma_2(\omega)}{\sigma_0(\omega)} = r_d^2 (1 + B\omega^2)$, $B \propto \gamma^{-2}$.
On dimensional grounds, $\gamma$ should be the characteristic frequency scale, so this makes intuitive sense.

\end{document}